# Dynamical effects of subducting ridges: Insights from 3-D laboratory models

J. Martinod[(1)], F. Funiciello[(2)], C. Faccenna[(2)], S. Labanieh[(1)], & V. Regard[(1)].

[(1)] LMTG, 14 avenue E. Belin, Université Paul Sabatier, 31400 Toulouse, France

[(2)] Dipartimento di Scienze Geologiche, Università degli Studi “Roma TRE”, Roma, Italy

J. Martinod : martinod@lmtg.obs-mip.fr; tel. (+33) 5 61 33 26 66; fax. (+33) 5 61 33 25 60

## Abstract

We model using analogue experiments the subduction of buoyant ridges and plateaus to study their effect on slab dynamics. Experiments show that simple local (1D) isostatic considerations are not appropriate to predict slab behaviour during the subduction of a buoyant ridge perpendicular to the trench, because the rigidity of the plate forces the ridge to subduct with the dense oceanic lithosphere. Oceanic ridges parallel to the trench have a stronger effect on the process of subduction because they simultaneously affect a longer trench segment. Large buoyant slab segments sink more slowly into the asthenosphere, and their subduction result in a diminution of the velocity of subduction of the plate. We observe a steeping of the slab below those buoyant anomalies, resulting in smaller radius of curvature of the slab, that augments the energy dissipated in folding the plate and further diminishes the velocity of subduction. When the 3D geometry of a buoyant plateau is modelled, the dip of the slab above the plateau decreases, as a result of the larger velocity of subduction of the dense “normal” oceanic plate on both sides of the plateau. Such a perturbation of the dip of the slab maintains long time after the plateau has been entirely incorporated into the subduction zone. We compare experiments with the present-day subduction zone below South America. Experiments suggest that a modest ridge perpendicular to the trench such as the present-day Juan Fernandez ridge is not buoyant enough to modify the slab geometry. Already subducted buoyant anomalies within the oceanic plate, in contrast, may be responsible for some aspects of the present-day geometry of the Nazca slab at depth.

## 1. Introduction

The oceanic seafloor is densely populated by so called “aseismic ridges”. They are areas of overthickened oceanic crust (seamount chains, island arcs, plateaus) isostatically more buoyant than normal oceanic lithosphere (Vogt, 1973; Cloos, 1993). An estimated 10.000 mapped aseismic ridges are present in the Pacific only (Kirby, personal communication). Globally, many aseismic ridges are actively subducting, such as for instance the Carnegie, Nazca and Juan Fernandez ridges below South America, the Entrecasteaux ridge below Vanuatu, or the Louisville ridge below the Tonga-Kermadec arc (Vogt, 1973; McCann & Sykes 1984; Collot et al., 1985; Cloos, 1993; Gutscher et al, 2000; Ruellan et al., 2003). What are the fate and the effects of these subducting objects? It has long been observed that the subduction of these features leads to “uncommon” subduction. It has been suggested that the subduction process as well as the geometry of plate boundary can be strongly affected by subduction of aseismic ridges because of the associated buoyant force (Vogt, 1973; Vogt *et al.*, 1976; Kelleher & McCann, 1977). The more evident effects of the subduction of aseismic ridges include apparent change in deep and shallow seismic activity (Vogt *et al.*, 1976; Kelleher & McCann, 1977; Nur & Ben Avraham, 1983; McCann & Sykes, 1984; Adameck *et al.*, 1987; Gutscher *et al.*, 2000), volcanic-arc segmentation and/or inhibition of arc volcanism (Nur & Ben Avraham, 1983; McGaery *et al.*, 1985.), coastal geomorphic modifications characterized by a rapid crustal uplift of the overriding plate (Chung & Kanamori, 1978; Gardner *et al.*, 1992; Macharé & Ortlieb, 1992; Taylor *et al.*, 1995). However, a proper classification of these effects applicable to all worldwide-subducted ridges is not yet available. This is mainly due to the fact that the character at the plate margin can widely change in response to the obliquity of the colliding structure with respect to the trench and in response to the speed of both incoming and overriding plate.

Previous studies relating to the effects produced by the subduction of aseismic ridges have included descriptive (Vogt, 1973, Kelleher & McCann, 1977, Nur & Ben Avraham, 1983, McCann & Habermann, 1989, Bouysse & Westercamp, 1990, Gutscher *et al.*, 2000), analytical (Chung & Kanamori, 1978) and numerical models (Kodama, 1984, Collot *et al.*, 1985, Geist *et al.*, 1993, van Hunen *et al.*, 2002). However, models quantifying effects of aseismic ridges on the dynamics of subduction in a fully 3-D setting are still lacking. With the present work we contribute to fill this gap.

We perform 3-D analogue models simulating the subduction of an oceanic plate into the mantle to study whether and how density heterogeneities within the subducting plate may control the geometry as well as the dynamics of subduction. In those experiments, plates contain buoyant regions to simulate the presence of oceanic ridges or plateaus entering into subduction. Our models show that the popular idea, typified by the Mariana example, that old plates should subduct vertically into the mantle is not adequate, as dense slabs are prone to retreat rapidly, favouring a shallowing of the slab. Moreover, we show that 1D buoyancy (local isostatic) considerations are not appropriate to predict the slab attitude. The effect of the subduction of buoyant material and its tendency to flatten upward the slab is counterbalanced by the negative pull of the surrounding lithosphere and, hence, its effect is very much dependent on the strength and coupling between ridges and slab itself. Our results also point out the difficulties to obtain a flat-like subduction as described below Central Peru (3°-15S) and North-Central Chile (27°-33°S).

## 2. Experimental set-up

*Boundary conditions and experimental geometry*

The experimental set-up adopted here is similar to that used by Funiciello *et al.* (2004). We model the upper mantle asthenosphere, between the base of the lithosphere and the 660 km discontinuity, using honey. Honey fills a rigid Plexiglas tank. In the centre of the tank, a silicone plate is deposited above honey to simulate the oceanic lithosphere before subduction (Figure 1). The silicone plate is negatively buoyant, and it sinks within the honey under its own weight simulating the subduction of a dense oceanic lithosphere.

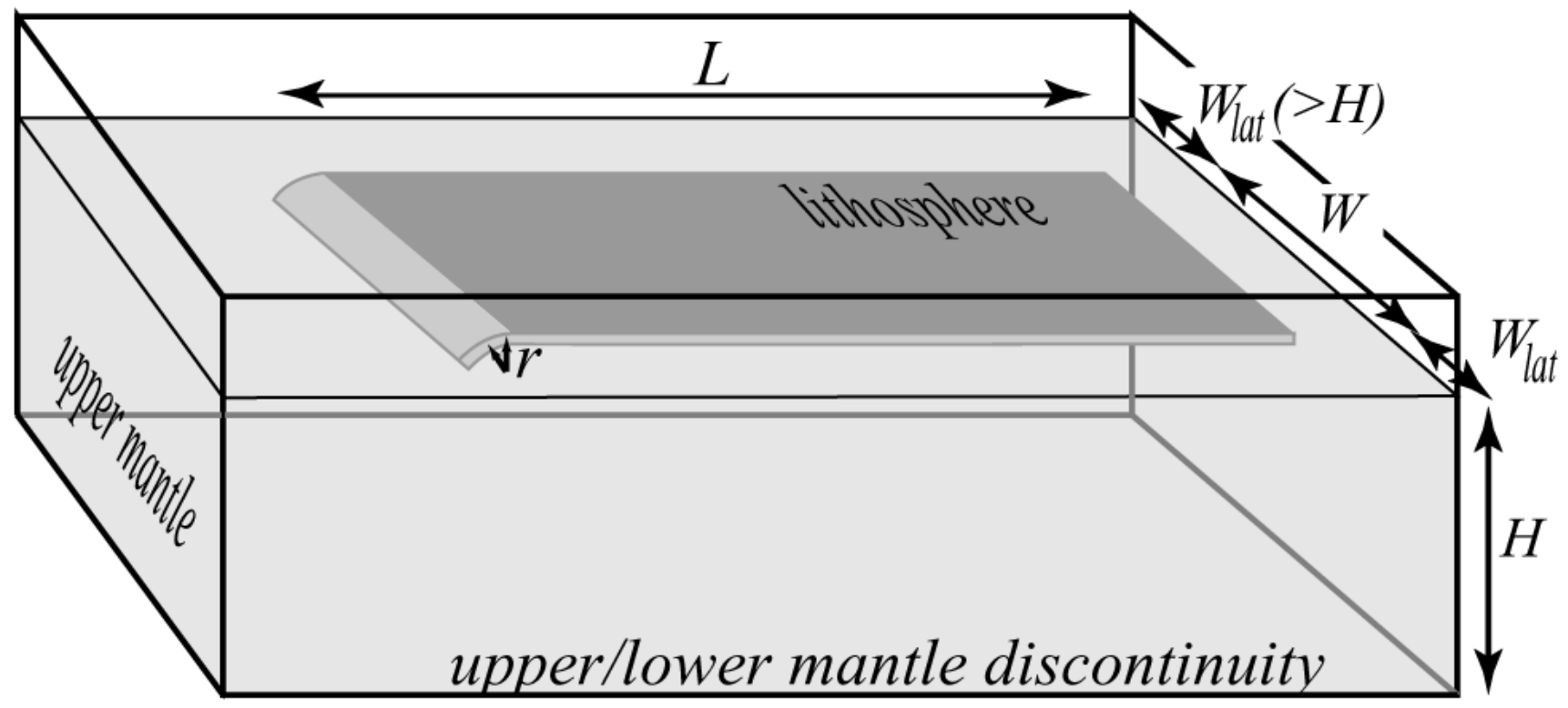

**Figure 1: 3-D view of the experimental set-up. The silicone layer modelling the oceanic lithosphere initially floats above the dense syrup modelling the upper mantle. The bottom of the Plexiglas tank represents the high-gradient viscosity increase at the lower mantle - upper mantle boundary. The subduction of the silicone plate is manually initiated at the beginning of experiment.**

We neglect the role of thermal convection in the subduction process, assuming that the upper mantle behaves as a passive Newtonian fluid and that the circulation within it is only caused by the plate/slab system. The reference frame of these experiments is the box boundaries. It can be considered as the experimental analogue of the fixed hot spot reference frame. We simply simulate the 660 km upper mantle-lower mantle discontinuity as an impermeable barrier (the bottom of the box). The validity of this approximation is confirmed by previous studies which found that the direct penetration of the slab through the transition zone is inhibited if the time-scale of the analysed process is limited (order of few tens of million years) and if the viscosity increase at the upper-lower mantle boundary is at least one order of magnitude (Davies, 1995; Guillou-Frottier *et al.*, 1995; Christensen, 1996; Funiciello *et al.*, 2003).

The overriding plate is not modelled in order to keep the set-up as simple as possible. Doing so, we assume that: (1) The fault zone between subducting and overriding plates is weak (Zhong & Gurnis, 1994; Tichelaar & Ruff, 1993; Conrad & Hager, 1999); (2) The velocity of the overriding plate is negligible with respect to the velocity of the trench. Therefore the applicability field of our experimental results is limited to natural cases characterized by this specific condition.

The distance between the plate and box sides ($W_{lat}$) is greater than the size of the advected cell within the upper mantle, which in first approximation corresponds to the thickness $H$ of the advecting layer modeling the upper mantle (honey layer) (see figure 1). Hence, during experiments, the process of subduction is not affected by the lateral confined box boundaries and can be considered as a proper analogue of a laterally unconstrained natural system, i.e. a system in which the upper mantle easily bypasses laterally the slab in case the slab is moving with respect to the mantle flow (Funiciello *et al.*, 2004).

The lithospheric plate is free to move responding self-consistently to subduction kinematics ("free ridge" sensu Kincaid & Olson, 1987). In fact, we assume that plates are completely surrounded by weak fault zones (trench and transform faults) whose equivalent viscosity is that of the upper mantle. These conditions result in a faster velocity of subduction (King & Hager, 1990) and ensure the maximum mobility of the plate.

*Rheology of the oceanic lithosphere*

We select silicone putty (Rhodrosil Gomme, PBDMS + iron fillers) to model lithospheric plates. Silicone putty is a visco-elastic material but the elastic component is negligible for the applied experimental strain rate. Hence, it can be considered as a quasi-Newtonian fluid in which stress increases linearly with strain rate (Weijermars, 1986). Therefore, we assume, following the experimental choice adopted by previous authors (Kincaid & Olson, 1987; Griffiths & Turner, 1988; Griffiths *et al.*, 1995; Guillou-Frottier *et al.*, 1995, Faccenna *et al.*, 1999), that the slab can be modelled as a Newtonian viscous rather than plastic (Shemenda, 1993) plate. This assumption is justified by the still rather elusive knowledge of the slab/lithosphere rheology (Karato *et al.*, 2001), and because the lithosphere in first approximation behaves like a viscous fluid in processes corresponding to global tectonics time scales (Tao & O'Connell, 1993). We further simplify the slab rheology using a Newtonian fluid whereas laboratory data indicates that upper mantle materials should obey to a creep power law of deformation (Brace & Kohlstedt, 1980). The ratio of the oceanic lithosphere viscosity over the upper mantle viscosity is set to roughly $10^4$ (Table 1), which corresponds to a reasonable possible value given our poor knowledge of the mantle rocks rheology (Davies & Richards, 1992; Mitrovica & Forte, 1997).

| | PARAMETER | NATURE | REFERENCE MODEL |
|---|---|---|---|
| g | Gravitational acceleration ($m/s^2$) | 9.81 | 9.81 |
| | Thickness (m) | | |
| hl | Oceanic lithosphere | 80000 | 0.012 |
| H | Upper mantle | 670000 | 0.1 |
| | *Scale factor for length* | $L_{model}/L_{nature} = 1.5\ 10^{-7}$ | |
| | Buoyancy ($kg/m^3$) | | |
| $\Delta\rho_L$ | "normal" 50 My old oceanic lithosphere | -35 | -56 |
| $\Delta\rho_{Pl}$ | 50 My-old ocean, 1800 m high plateau | +16 | +25 |
| | *Buoyancy ratio* | $(\Delta\rho_L)_{model} / (\Delta\rho_L)_{nature} = 1.6$ | |
| | Viscosity (Pa.s) | | |
| $\eta_l$ | Oceanic lithosphere | $\sim 10^{24}$ | $2.7\ 10^5$ |
| $\eta_{um}$ | Upper mantle | $\sim 10^{20}$ | 27 |
| $\eta_{lm}$ | Lower mantle | $\sim 10^{22}$ | $\infty$ |
| | *Scale factor for viscosity* | $\eta_{model} / \eta_{nature} \sim 2.7\ 10^{-19}$ | |
| t | Characteristic time (s) | $3.1\ 10^{13}$ | 35 |
| | $(t_{nature}/t_{model} = (\Delta\rho gh)_{model} / (\Delta\rho gh)_{nature} \cdot (\eta_{nature} / \eta_{model}))$ | (1 My) | |
| U | Characteristic velocity (m/s) | $3.16\ 10^{-9}$ | $4.2\ 10^{-4}$ |
| | $(U_{nature} / U_{model}) = (L_{nature} \cdot t_{model}) / (L_{model} \cdot t_{nature})$ | (10 cm/y) | (0.42 mm/s) |

**Table 1. Scaling of parameters in nature and in laboratory for a reference experiment.**

*Buoyancy of the oceanic plate*

In this paper, we focus on the effect of oceanic ridges on the process of subduction. Thick oceanic crust is responsible for the submarine relief of oceanic ridges and plateaus. A thicker crust increases the average buoyancy of the lithospheric plate, and the oceanic plate may become lighter than the asthenosphere, depending on the age of the plate and on the thickness of the crustal layer (Molnar & Gray, 1979; Cloos, 1993).

In the following, we model the subduction of a 80 km-thick lithosphere (~50 My-old oceanic plate). In case the thickness of the oceanic crust is "normal" (~7 km), the 50 My-old oceanic lithosphere is negatively buoyant with respect to the asthenosphere, and its average buoyancy $(\Delta\rho_L)_{Nature}$ = –35 kg/m$^3$ (Cloos, 1993; Gutscher et al., 2000). The same plate becomes neutrally buoyant if the thickness of the basaltic crust increases to 14 km. Local isostatic equilibrium implies that such a plateau corresponds to a 1250 m local relief above the "normal crust" oceanic floor. Higher plateaus and ridges above 50 My-old oceanic plates correspond to positively buoyant lithospheric segments.

| **Experiment** | | **1** | **2** | **3** | **4** | **5** | **6** | **7** | **8** | **9** | **10** |
|---|---|---|---|---|---|---|---|---|---|---|---|
| Initial set-up (see Fig. 2) | | (a) | (a) | (b) | (c) | (c) | (d) | (d) | (d) | (e) | (e) |
| L | | 400 | 500 | 400 | 510 | 500 | 400 | 400 | 500 | 500 | 570 |
| W | | 200 | 400 | 200 | 200 | 200 | 200 | 200 | 400 | 400 | 400 |
| Thickness silicone plate (h) | | 11 | 13 | 12 | 13 | 10 | 11.5 | 12 | 13 | 14 | 12 |
| l | | - | - | 200 | 300 | 300 | 200 | 200 | 300 | 300 | 200 |
| Lr | | - | - | 200 | 50 | 30 | 200 | 200 | 200 | 50 | 50 |
| Wr | | - | - | - | - | - | 30 | 30 | 30 | 150 | 150 |
| Thickness honey (H) | | 104 | 100 | 104 | 103 | 103 | 104 | 104 | 100 | 100 | 100 |
| Heavy silicone | Density | 1497 | 1497 | 1497 | 1497 | 1497 | 1497 | 1497 | 1497 | 1497 | 1497 |
| | Viscosity | $2.7x10^5$ | $2.7x10^5$ | $2.7x10^5$ | $2.7x10^5$ | $2.7x10^5$ | $2.7x10^5$ | $2.7x10^5$ | $2.7x10^5$ | $2.7x10^5$ | $2.7x10^5$ |
| Light silicone | Density | - | - | 1416 | 1416 | 1416 | 1416 | 1416 | 1416 | 1416 | 1426 |
| | Viscosity | - | - | $2.6x10^5$ | $2.6x10^5$ | $2.6x10^5$ | $2.6x10^5$ | $2.6x10^5$ | $2.6x10^5$ | $2.6x10^5$ | $2.6x10^5$ |
| honey | Density | 1422 | 1441 | 1422 | 1422 | 1422 | 1422 | 1422 | 1441 | 1441 | 1427 |
| | Viscosity | 12 | 27 | 12 | 12 | 12 | 12 | 12 | 27 | 27 | 13 |
| Relative plateau elevation in Nature (m) | | - | - | 1360 | 1360 | 1360 | 1360 | 1360 | 1800 | 1800 | 1250 |

**Table 2. Experimental parameters. The label of experiments, from (a) to (e), refer to the different geometries of the light area within the silicone plate (see Fig. 2). See also Fig. 2 for the description of *W, L, l, w, Wr,* and *Lr*. Length are given in mm, viscosities in Pa.s, and densities in kg.m$^{-3}$.**

In the experiments presented below, the density of the plate modelling the oceanic lithosphere is set to correspond to the average density of the lithospheric plate. To model oceanic ridges in which the thicker crust increases the buoyancy of the lithosphere, we diminish the density of the silicone plate over its whole thickness. Such an approximation should not modify the general behaviour of the subduction zone, assuming the oceanic crust is not delaminated from the mantle during the process of subduction. We also consider that the thickness of the lithosphere below oceanic ridges is similar to that of the neighbouring oceanic plate.

Buoyancy values and the scaling of density contrasts are detailed in Tables 1 and 2. In experiment 8, for instance, the buoyancy of the silicone plate modelling the "normal" oceanic plate $(\Delta\rho_L)_{Model}$ = –56 kg/m$^3$, while the buoyancy of the plate modelling the plateau lithosphere $(\Delta\rho_{Pl})_{Model}$ = +25 kg/m$^3$ (Table 1). If we consider we are modelling a 50 My-old oceanic lithosphere, the buoyancy scaling factor $(\Delta\rho_L)_{Model}$ / $(\Delta\rho_L)_{Nature}$ = 1.6, and the corresponding buoyancy of the light plate in nature ($(\Delta\rho_{Pl})_{Nature}$ = +16 kg/m$^3$) represents that of a lithosphere with a 17 km-thick crust. This approximately corresponds to a ridge or plateau 1800 m higher than the "normal" 50 my-old oceanic plate. We calculated the relative elevation of plateaus modelled in the experiments presented below (Table 2). It varies between 1250 and 1800 meters.

### *Isothermal experiments*

We are forced by laboratory limitations to neglect thermal effects during the process of subduction. These models represent an end-member essentially governed by the negative buoyancy of the slab. Temperature contrasts are modelled by density contrasts, and they stay constant throughout the experiments. In this view the slab is thought to be in a quasi-adiabatic condition. The high velocity of subduction (larger than an equivalent of 1 cm/yr) recorded in our experiments justifies this assumption ensuring that advection overcomes conduction. These isothermal experiments also imply the impossibility to consider the fundamental role of phase changes in slab dynamics (Christensen & Yuen, 1984; Pysklywec & Mitrovica, 1998). In particular, we simulate the impediment of the slab to penetrate directly into the lower mantle to result only from the increase of viscosity with depth, this approximation being in agreement with the results of previous simulations where the effect of a viscosity increase with depth overcomes the one exerted by phase transformations (Bunge *et al.*, 1997; Lithgow-Bertelloni & Richards, 1998).

### *Experimental procedure*

Materials have been selected to respect the standard scaling procedure for stress scaled down for length, buoyancy, viscosity, in a natural gravity field ($g_{model}$=$g_{nature}$) as described by Weijermars & Schmeling (1986) and Davy & Cobbold (1991). Experimental parameters and the scaling relationships of a reference experiment are listed in Table 1. From the simple scaling equations of Table 1 we find that 35 seconds and 1 centimetre in the experiment correspond to 1 Myr and 70 km in nature, respectively.

In the initial configuration, the leading edge of the silicone plate is forced downwards to a depth of 3 cm (corresponding to about 200 km in nature) inside honey as a means to start subduction process. Problems linked to initiation of subduction (e.g. Faccenna *et al.*, 1999; Regenauer-Lieb *et al.*, 2001) are beyond the scope of this paper. Each experiment is monitored with a sequence of photographs taken in regular time in lateral and top views. We measure from them the motion of both the plate and the trench in time (their algebraic sum being equal to the amount of subduction) and the dip of the slab.

## 3. Experimental results

Experiments are designed as a parameter-searching test to analyse the role played by the density of the plate on the geometry of the slab and on the dynamics of subduction. We first study the subduction of a homogeneous dense plate ("reference experiment"). Then, we look at the subduction of a light plate following an oceanic subduction. Afterwards, we include density anomalies that model the presence of oceanic plateaus or ridges (Figure 2) and we change in a systematic way their orientation and length. Experiments with different plate width, *W*, have been performed. We observe that the process of subduction develops very similarly in both set of experiments, and we present below indifferently narrow or large experiments to illustrate the effect of density contrasts on subduction (see Bellahsen *et al.* (2005) for a systematic experimental study of the influence of the plate width on the subduction process).

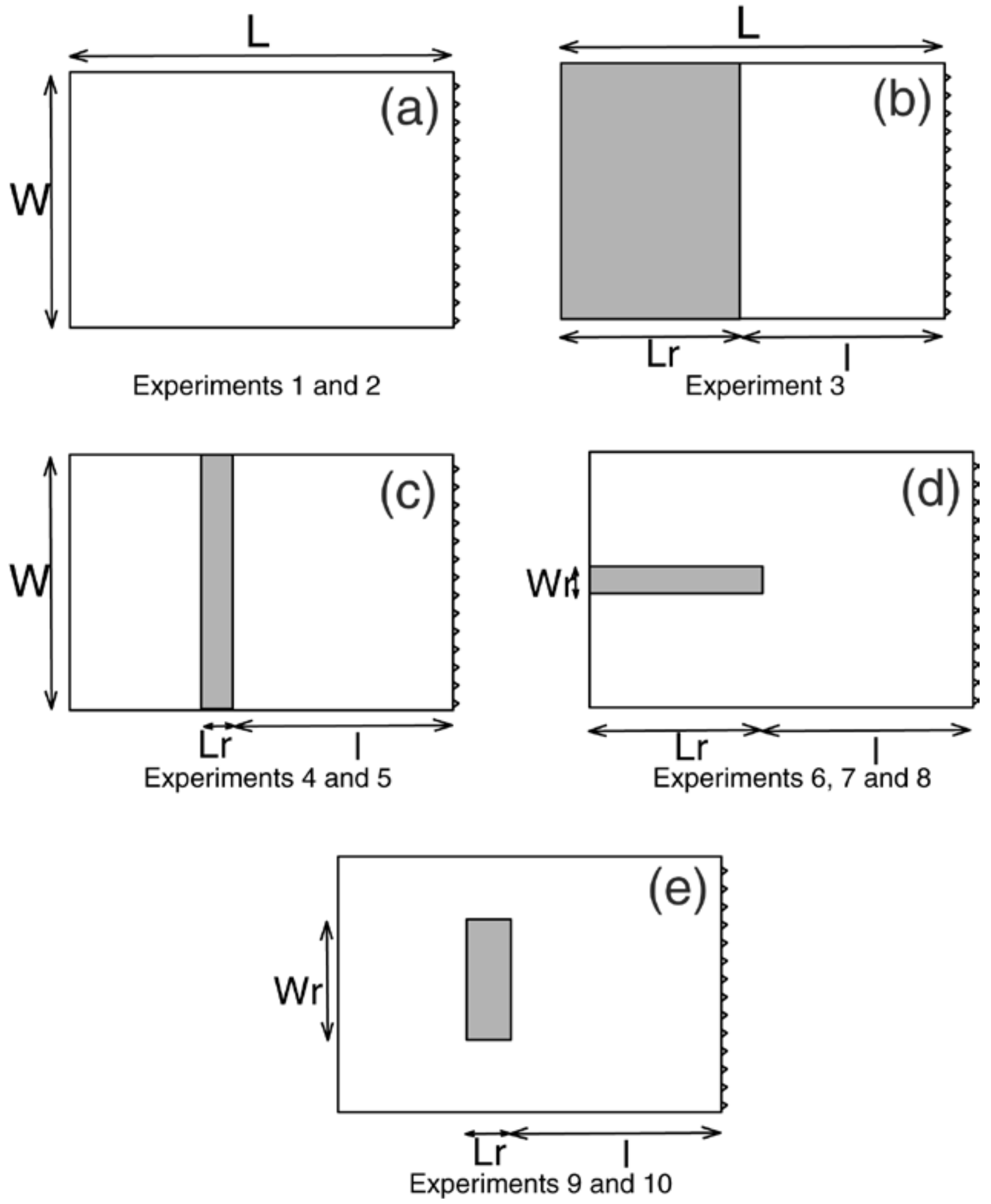

**Figure 2: Top view of the different configurations of the subducting plates used in the present study. White areas correspond to dense silicone modelling a heavy oceanic lithosphere, grey areas to lighter silicone modelling buoyant oceanic lithosphere (oceanic ridge and/or plateau). Subduction initiates on the right side of the plates. (a) Subduction of a homogeneous oceanic plate ("Reference oceanic experiment"). (b) Oceanic subduction followed by subduction of a buoyant plate (c) oceanic ridge parallel to the trench. (d) Oceanic ridge perpendicular to the trench. (e) Oceanic plateau. See Table 2 for details.**

### 3.1 Oceanic subduction (Reference experiments: experiments 1 and 2; Fig. 2a)

The behaviour of similar experiments has already been described in Funiciello *et al.* (2004) and Bellahsen *et al.* (2005). These authors show that the geometry and behaviour of the subduction zone can vary significantly using variable combinations of thickness, viscosities, densities of the plate and mantle (Bellahsen *et al.,* 2005) and depending on the possibility for the mantle to bypass the slab laterally ("opened" or "closed" configuration, Funiciello *et al.*, 2004). When they use geometrical configurations close to that we adopt in our reference experiments, Funiciello *et al.* (2004) and Bellahsen *et al.* (2005) obtain experimental results that are similar to those described below.

To initiate subduction, we fold downward the leading edge of the silicone plate to a depth of 3 cm inside the honey (Figure 3a). Then, the subduction pursues alone, essentially pulled by the negative buoyancy of the slab. We observe that the process of subduction can be divided in two main phases, a steady-state regime of subduction establishing after an initial transitory episode. Figures 3 and 4 show the evolution of experiment 1. Experiment 2 models the subduction of a larger homogeneous oceanic plate (Table 2). Its evolution, essentially comparable to that of experiment 1, also reproduces the successive episodes of the subduction described below.

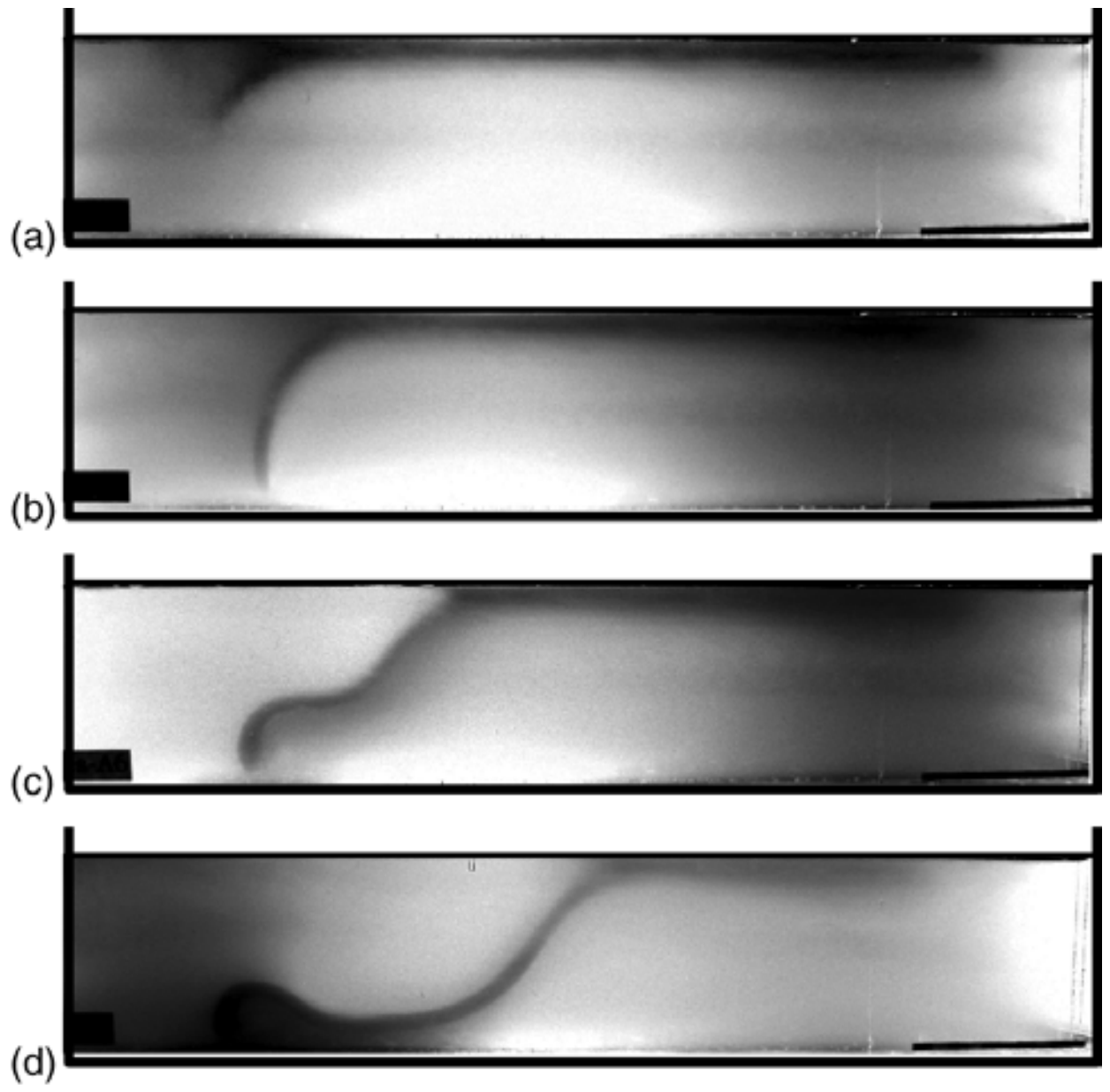

**Figure 3: Lateral view of experiment 1. (a) Beginning of the experiment. Subduction is starting. (b) After 35 seconds of experiment, the slab touches the bottom of the box. (c) After 4 minutes of experiment, the slab is falling to lie on the box. (d) After 7 minutes of experiment, subduction reaches a steady-state regime.**

<u>*Transitory regime of subduction*</u>

At the beginning of experiment, subduction accelerates and the slab dip increases up to a maximum angle close to 90° (Figure 3b). The subduction velocity increase results from the growth of the slab and the associated increase in slab pull force (Becker *et al.*, 1999). During this phase of the experiment, the oceanic plate moves rapidly toward the trench, this motion participating significantly to the velocity of subduction (Figure 4). The bottom of the tank is reached after 40 seconds of experiment. At that moment, the velocity of subduction diminishes significantly while the tip of the slab folds and deforms at depth (Figures 3c and 4). This transitional situation pursues for a few minutes, during which the absolute motion of the plate toward the trench is stopped, and the subducted tip of the slab falls down to lie horizontally on the bottom of the tank.

**Figure 4: Amount of subducted plate (continuous line) and absolute motion of the plate (dashed line) vs. time in experiment 1. The difference between the two lines corresponds to trench retreat. The vertical dotted line at 40 seconds marks the moment the tip of the slab attains the bottom of the box. The second dotted line at 5 minutes 20 seconds marks the moment the slab has fallen to lie horizontally above the bottom of the box. It is the beginning of the steady-state subduction. Black arrows mark the moment of the four photos in Fig. 3.**

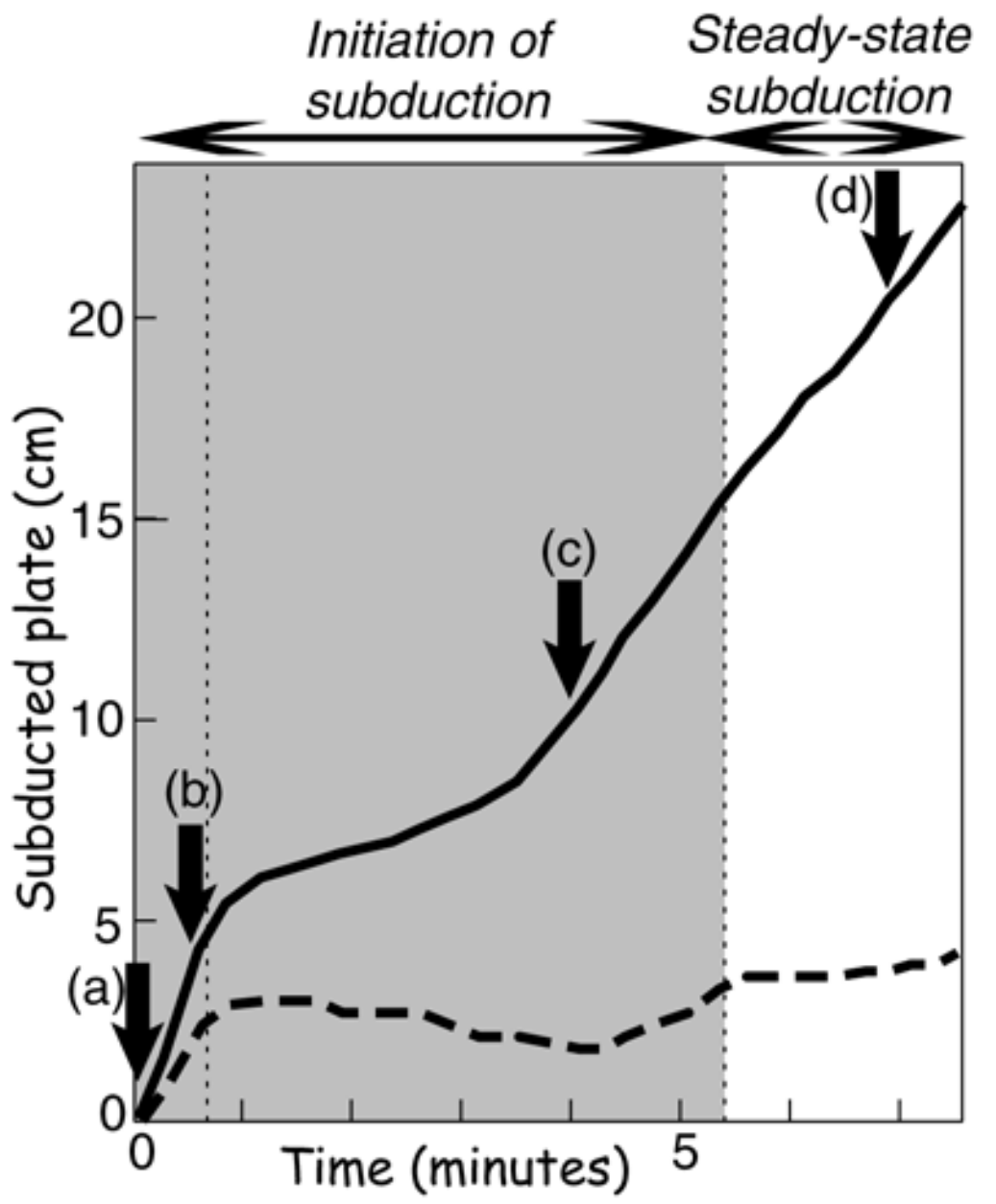

<u>*Steady-state regime of subduction*</u>

After the transitory phase, a steady-state regime establishes, during which the subduction velocity stabilizes and the slab dip maintains a constant value (Figure 4, see also e.g. Funiciello *et al.*, 2003). The constant velocity of subduction is the consequence of the constant value of the slab-pull force, since the part of the slab lying above the tank bottom does not exert any traction on the upper part of the plate. From that time, subduction essentially results, in those experiments, from the retrograde displacement of the trench toward the ocean, this process being always associated with significant displacements of the mantle on both sides of the retreating slab. The steady-state velocity of oceanic subduction is roughly 0.58 mm/s in experiment 1 with a dip of 50° (Figs 3d and 4, Table 3).

| experiment | time | $F_{SP}$ (N/m) | Radius of curvature (m) | Velocity of subduction ($x10^{-4}$m/s) | D |
|---|---|---|---|---|---|
| 1 | 350 | 0.97 | 0.056 | 5.8 | 2.45 |
| 2 | 1130 | 0.73 | 0.055 | 2.2 | 2.15 |
| 3 | 345 | 0.94 | 0.057 | 6.4 | 3.43 |
| 3 | 800 | 0.31 | 0.032 | 0.2 | 1.84 |
| 4 | 520 | 1.1 | 0.066 | 6.3 | 2.36 |
| 4 | 1030 | 0.48 | 0.041 | 0.7 | 2.51 |
| 5 | 400 | 0.94 | 0.057 | 6.0 | 1.86 |
| 5 | 630 | 0.56 | 0.044 | 2.6 | 2.94 |

**Table 3. Velocity of subduction in type (a), (b) and (c) experiments (see Fig. 2). The velocity of subduction $U$ obeys equation (2): $U = D F_{SP} r^3 / (2 \eta_L h^3)$, where $F_{SP}$ is the slab pull force, $\eta_L$ is the viscosity of the plate, $h$ is its thickness, $r$ is the radius of curvature of the slab, and $D$ is a constant whose value is approximately 2.5.**

The adopted experimental conditions result in the formation of a “mode 1” subduction, as defined by Bellahsen *et al.* (2005). This mode is characterized, following the initiation of the process, by the appearance of the steady-state regime described above, during which subduction essentially results from slab roll-back. During steady-state subduction, the velocity of subduction and the dip of the slab maintain constant. Therefore, this mode of subduction is ideal to evidence the dynamic changes that result from the subduction of buoyant anomalies. In the following experiments, buoyant anomalies always reach the trench after the steady-state regime has established. Results of these experiments should apply directly to natural retreating subduction zones, although the general qualitative conclusions we deduce from the analysis of these experiments should also be valid for other modes of subduction.

### 3.2 Subduction of a buoyant plate following an oceanic subduction (experiment 3; Fig. 2b)

Experiment 3 corresponds to the subduction of a plate, half of which is composed by a dense negatively buoyant silicone layer modelling a “normal” 50 Ma-old oceanic lithosphere, while the remaining half corresponds to a lighter positively buoyant plate (Table 2). At the beginning of experiment, the oceanic part of the plate subducts, as observed in experiment 1 (Figures 5 and 6), and after approximately 5 minutes, the slab falls down horizontally on the bottom of the tank. After 5mn 45s of experiment, the buoyant part of the plate enters into subduction. From that moment, we observe that the velocity of subduction and of trench retreat decreases (Figure 6), producing a steeping of the slab that attains a backward reclined position (dip of 100°). As a consequence, the radius of curvature of the slab diminishes. After approximately 13 minutes of experiment, the subduction almost stops: Its velocity is only 0.02 mm/s, vs. 0.7 mm/s when the light plate arrived at trench (Table 3).

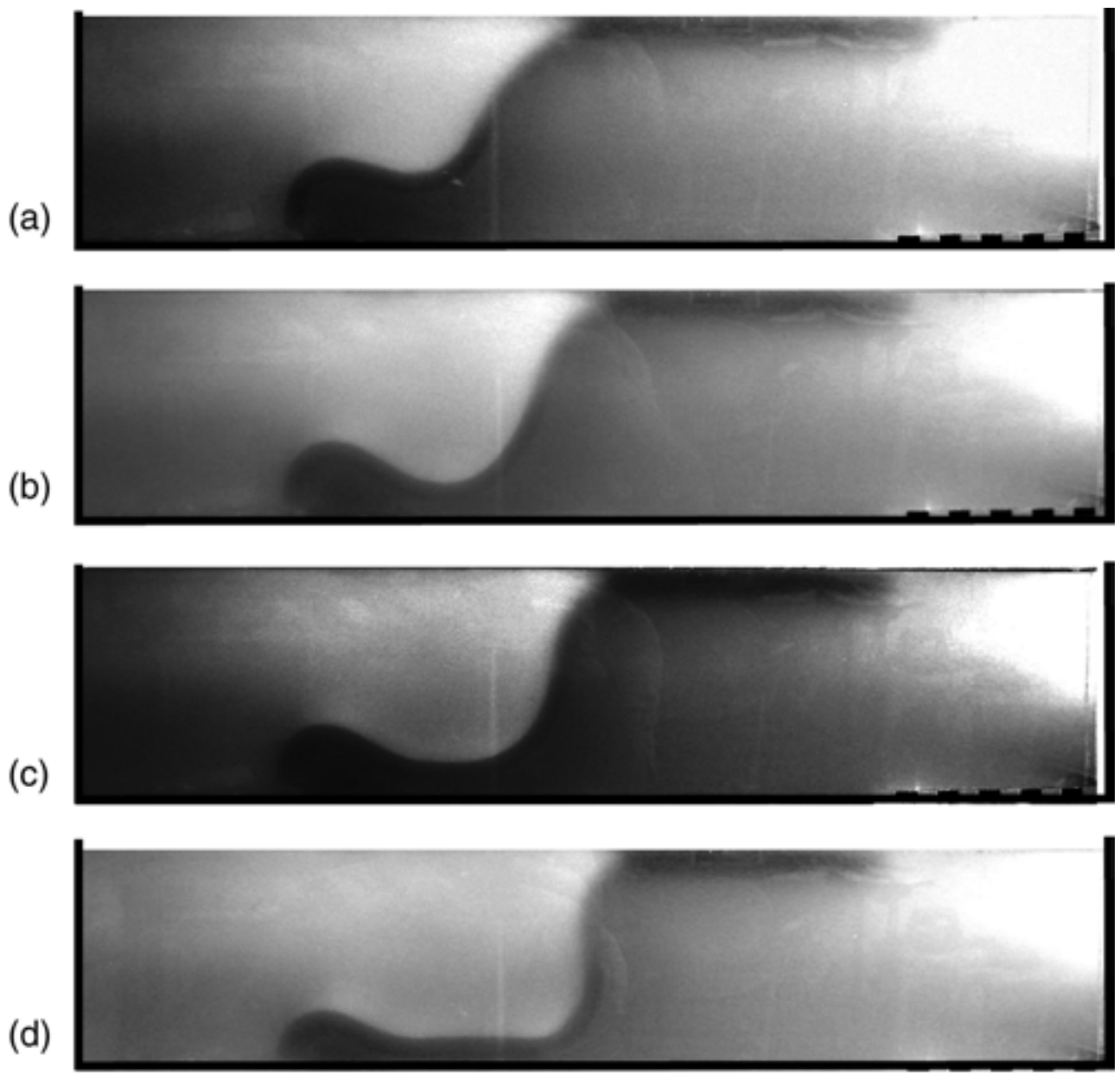

**Figure 5: Lateral views of experiment 3. (a) After 5mn 52s of experiment, the light part of the plate is arriving at trench. (b) After 7mn 30s of experiment. (c) After 9mn 10s of experiment. (d) After 13mn 15s of experiment. Subduction velocity has fallen close to zero. The slab is vertical, and its radius of curvature 2 times smaller than when the dense part of the plate was subducting.**

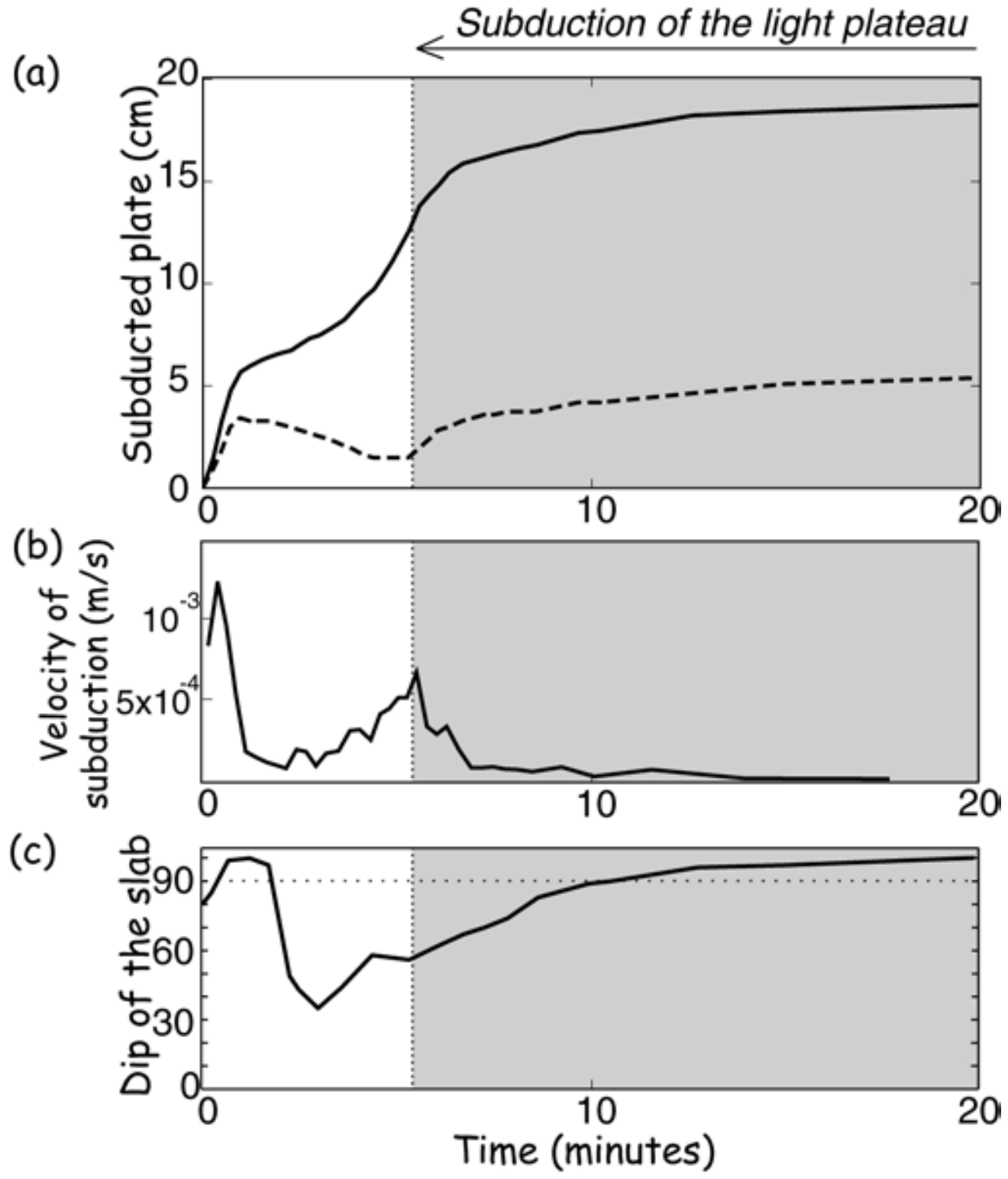

**Figure 6: (a) Amount of subducted plate (continuous line) and absolute motion of the plate (dashed line), (b) subduction velocity, and (c) dip of the slab, vs. time in experiment 3. The grey domain corresponds to the subduction of the light part of the plate. The entrance of the ridge at trench induces a sharp decrease of the velocity of subduction and an increase of the dip of the slab below the ridge. After 13 minutes of experiment, the subduction velocity is close to zero.**

### 3.3 Subduction of a ridge parallel to the trench (experiments 4 and 5; Fig. 2c)

Experiment 4 is set as the reference experiment (Experiment 1), except the presence of a 5 cm-large light silicone band that simulates a 330 km-large aseismic ridge resulting in an elevation of the sea floor of 1360 meters, parallel to the trench, and that crosses the plate. The extra topography perpendicular to the corresponding natural ridge would then be 4.5 $10^8$ $m^2$ (1360 x 330.$10^3$ $m^3$ per meter of ridge). This value can be compared, for instance with the extra topography of some natural ridges : the analysis of global relief ETOPO5 data show that the Nazca ridge, for instance, is 250 – 300 km large, and the associated extra topography perpendicular to the ridge varies between 2 $10^8$ and 4.5 $10^8$ $m^2$. The size of the Carnegie ridge is similar, the extra-topography associated to that ridge varying between 2 $10^8$ and 5 $10^8$ $m^2$. The Juan Fernandez ridge, in contrast, is much smaller, the associated extra-topography varying between 2 $10^8$ $m^2$ and zero. The ridge we model here is then comparable, although somewhat larger, to the Nazca ridge, and much larger than a modest ridge such as the present-day Juan Fernandez ridge.

After 6 minutes of experiment, the subducted slab deposits horizontally above the bottom of the Plexiglas tank, and the steady-state subduction regime occurs at the same velocity as in experiment 1 (Figures 7 to 9). The dip of the slab is roughly 50° as in the reference oceanic experiment. After 9 minutes of experiment, the light ridge reaches the subduction zone and begins to subduct within the mantle. The velocity of subduction and of trench retreat decreases to 12% of that of the subduction of the dense oceanic plate (0.7 vs. 6.3 mm/s), causing an increase of the dip of the slab below the ridge from 50° to approximately 90° (Figure 9). Above the ridge, the dip of the slab preserves smaller values of approximately 50°, i.e. similar to that of the slab in the reference oceanic experiment (Figure 8).

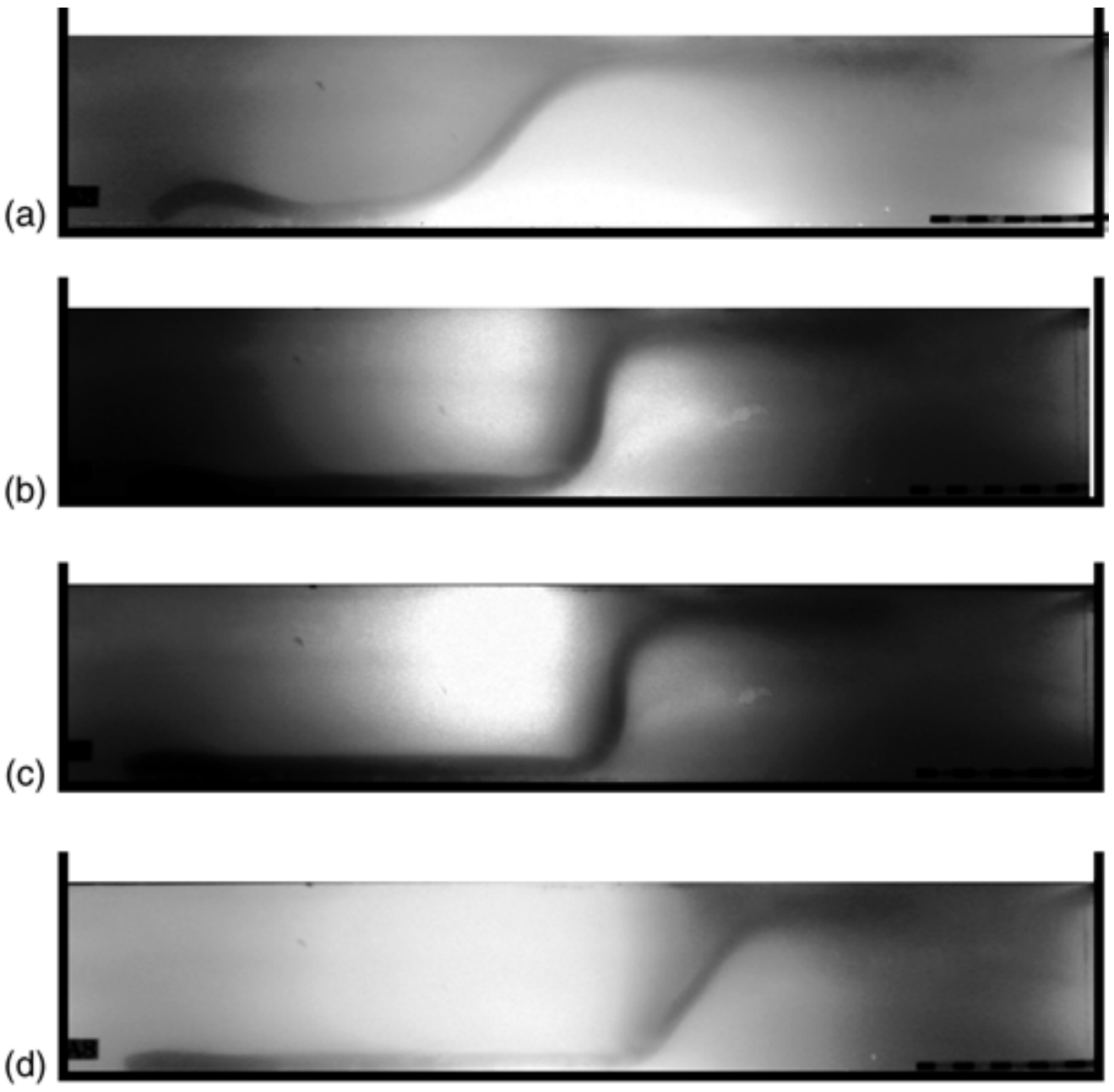

**Figure 7: Lateral views of experiment 4. (a) After 8mn 21s of experiment: steady-state subduction of a homogeneous dense oceanic plate. The dip of the slab is close to 50°. (b) After 13mn 12s of experiment, the light ridge is subducting. The velocity of subduction has dropped, the radius of curvature of the slab has diminished to roughly 4 cm, and the slab below the light ridge is nearly vertical. (c) After 17mn 17s of experiment, the entire light ridge is now subducted, but it maintains at narrow depth, and the dip of the slab below the ridge is still close to 90°. (d) After 25mn 14s of experiment, the ridge has now fallen within the deep upper mantle, subduction accelerates, and the dip of the slab is again close to 50°.**

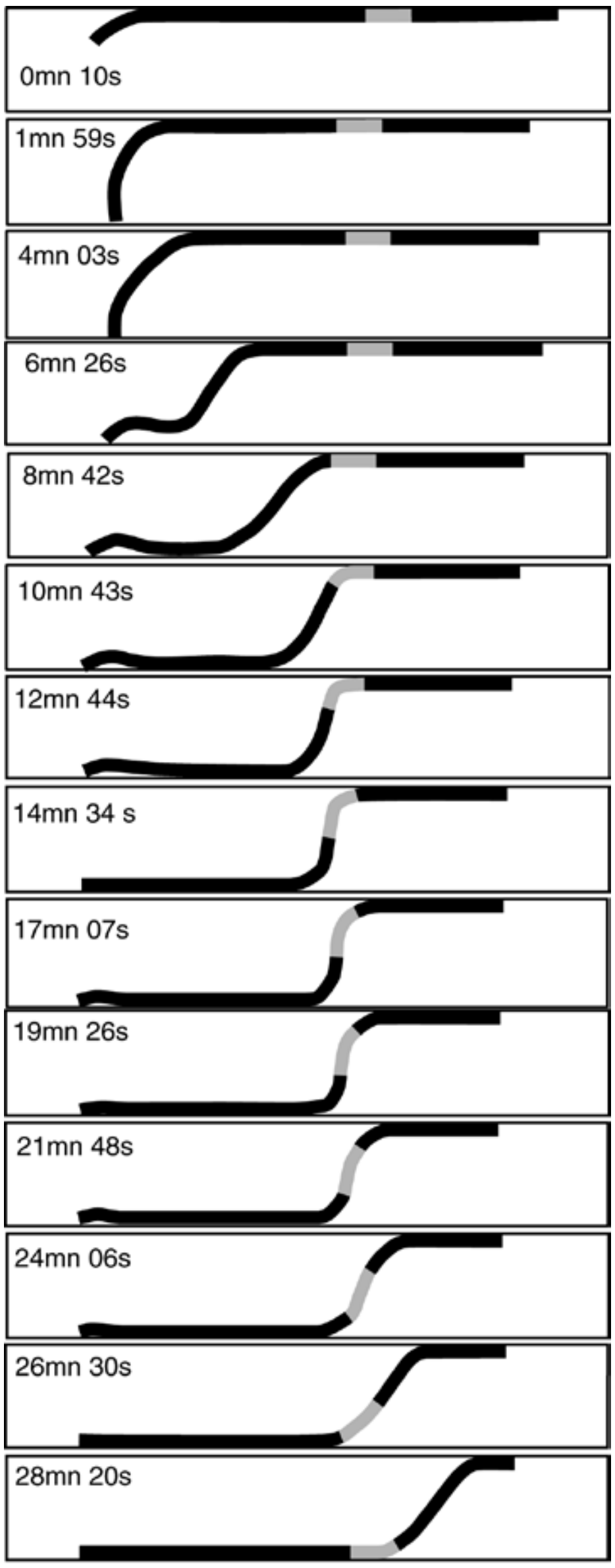

**Figure 8: Line drawing of lateral views of experiment 4. The light grey band marks the position of the light ridge within the plate. The subduction of the light ridge is marked by a diminution of the velocity of subduction and by the increase of the dip of the slab below the ridge.**

At the end of experiment, as the light ridge approaches the bottom of the tank, the dense subducted slab located above the ridge pulls again the lithospheric plate toward the subduction zone. Subduction velocity increases again, and the dip of the entire slab becomes again close to 50° (Fig. 8). Afterwards, the subduction of the oceanic plate pursues again, with the same velocity and geometry than before the subduction of the ridge (Figure 9).

**Figure 9: (a) Amount of subducted plate (continuous line) and absolute motion of the plate (dashed line), and (b) dip of the slab vs. time in experiment 4. The first vertical dotted line marks the moment the tip of the slab attains the bottom of the box. The second one corresponds to the end of the initiation of subduction. The grey band corresponds to the subduction of the light ridge parallel to the trench. The entrance of the ridge at trench induces a sharp decrease of the velocity of subduction and an increase of the dip of the slab below the ridge. Slow subduction continues as long as the light ridge is sinking within the upper mantle, and the rapid steady-state subduction only starts again after 28 minutes of experiment.**

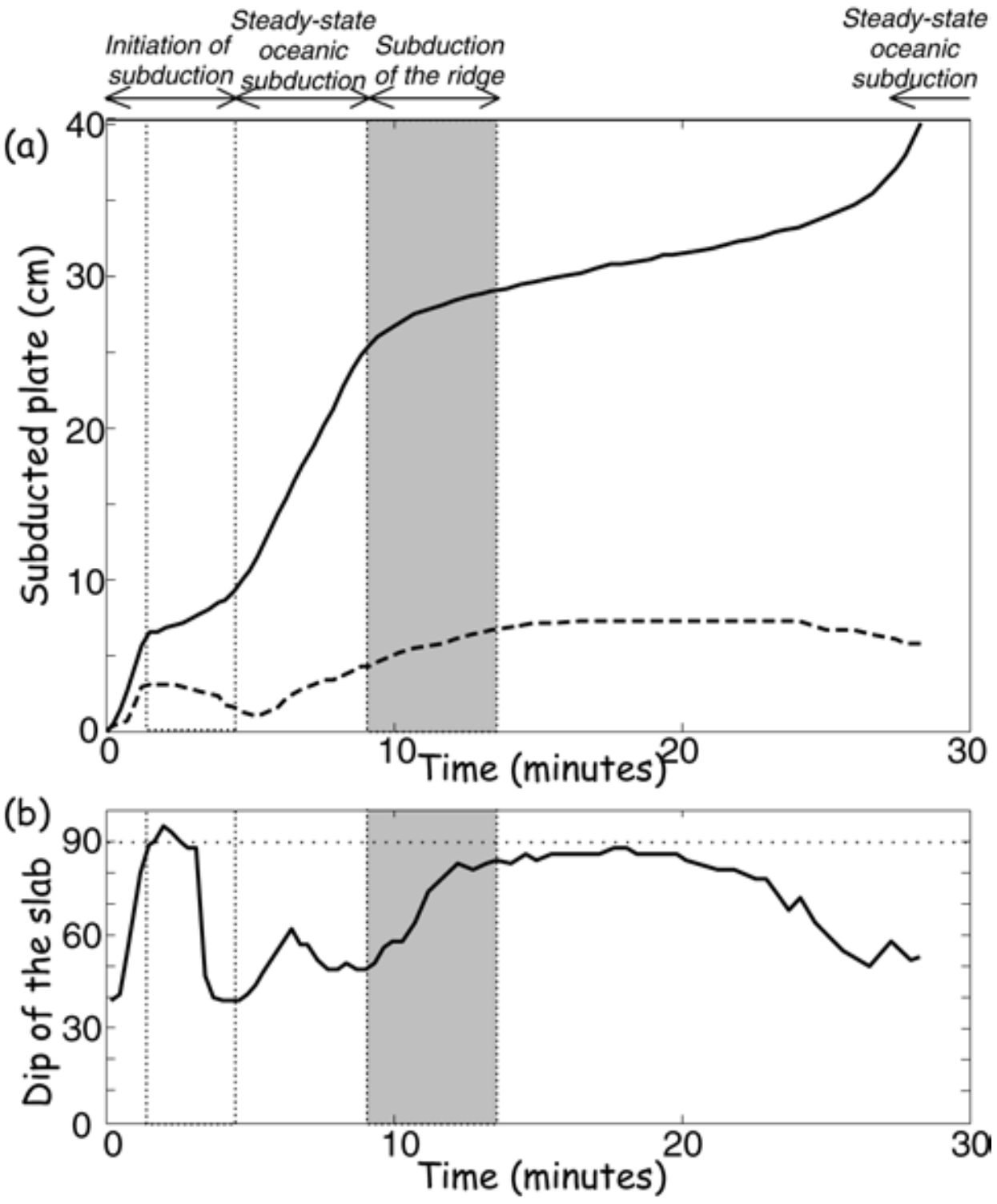

This experiment shows that the subduction of a light segment of lithosphere parallel to the trench has strong effects on the kinematics of subduction and on the geometry of the slab. The velocity of subduction and of slab retreat is reduced, producing an increase of the dip of the slab, which in turn results in a diminution of the radius of curvature of the slab (Figure 7).

Experiment 5 also models the subduction of a ridge parallel to the trench. The subducting ridge, however, is narrower than in experiment 4 (3 cm vs. 5 cm, see Table 2). This would correspond in nature to a ridge with a perpendicular to the ridge extra topography of ~2.8 $10^8$ $m^2$. Here again, we observe that the subduction of the ridge results in a sharp decrease of the velocity of subduction (from 0.6 to 0.26 mm/s). We also note an increase of the dip of the slab, from 54° during the subduction of the dense part of the plate, to 84° during the subduction of the ridge.

Note that in nature the subduction of an aseismic ridge parallel to the trench is a rather uncommon phenomenon. Ridges follow the direction of the absolute motion of the plate, and for that reason they usually subduct perpendicularly to the trench. However, subduction of oceanic ridges approximately parallel to the trench may occur, for instance in case the absolute motion of the subducting plate changed between the creation of the ridge and its subduction. This may have occurred during the Miocene, for instance, to the Juan Fernandez ridge subducting below South America (Yañez *et al.*, 2001).

### 3.4 Subduction of a ridge perpendicular to the trench (experiments 6, 7 and 8; Fig. 2d)

Experiment 8 is constituted of a 50x40 cm heavy silicone plate modelling the oceanic lithosphere. A 3 cm-large, 20 cm-long light band is placed perpendicularly to the trench to simulate a 200 km-large ridge (Table 2; Figures 2e and 10). The relative buoyancy of this ridge is approximately the same as that of experiment 5, and it may constitute a good analogue of the biggest ridges (Nazca and Carnegie) that subduct below south America. The initiation of subduction is similar to that observed in Experiment 1 (Fig. 11). Steady-state subduction, essentially governed by trench retreat, begins after 10 minutes of experiment, at a constant speed of 0.22 mm/s, as in experiment 2. The dip of the slab is approximately 60°, which also corresponds to the dip observed in experiment 2.

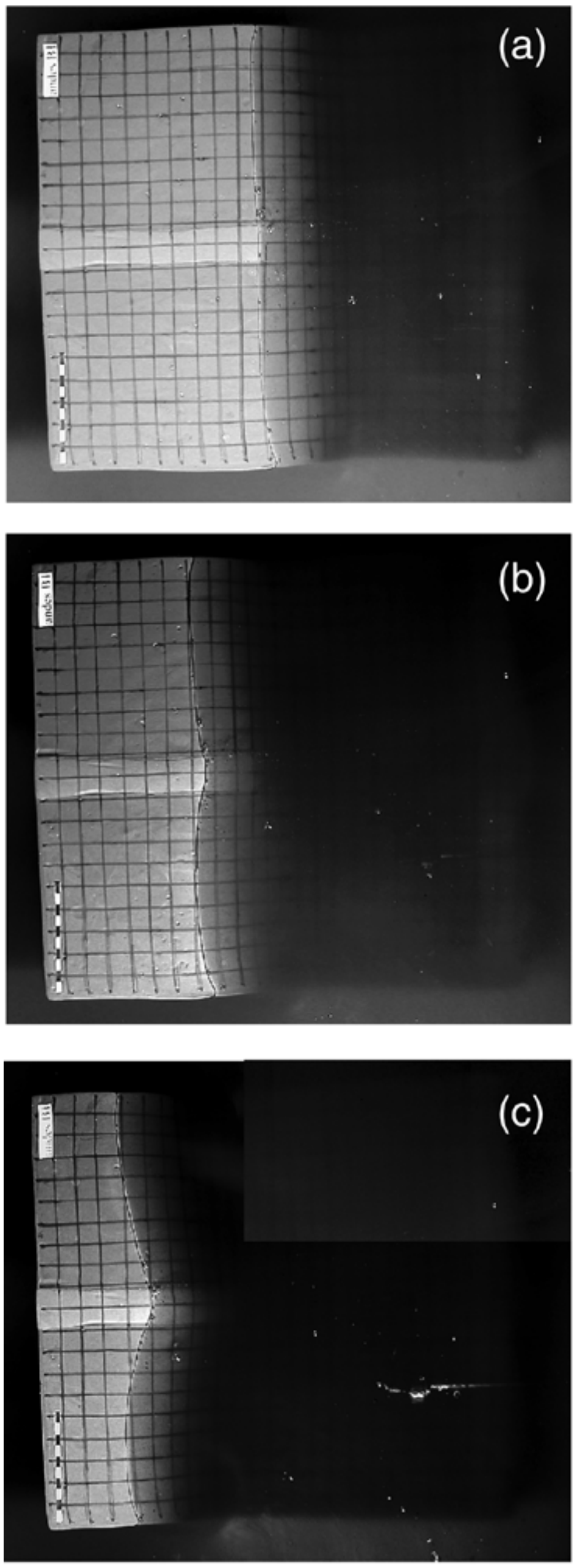

**Figure 10: Upper view of experiment 8. The 2 cm per 2 cm grid of passive markers gives the scale. (a) After 21mn12s of experiment, the ridge is arriving at trench. (b) After 25mn17s. (c) After 30mn40s, the geometry of the trench is influenced by the subduction of the ridge. The dip of the slab, however, has not changed.**

After 21 minutes, the tip of the ridge reaches the trench. The process of subduction, however, pursues without any major change, neither in its velocity, nor in the dip of the slab, during several minutes. After 25 minutes of experiment, the velocity of subduction in front of the ridge becomes slightly smaller (20% smaller) than that of the rest of the plate. The small difference in subduction velocity between the ridge and the rest of the plate produces an arched shape of the trench in front of the ridge, that slowly amplifies (Figure 10). The subduction velocity of the heavy plate is influenced by the presence of the subducting ridge at distances smaller than roughly 6 cm, equivalent to 420 km in nature. Far from the ridge, in contrast, we do not measure any significant change in the velocity of subduction (Figure 11).

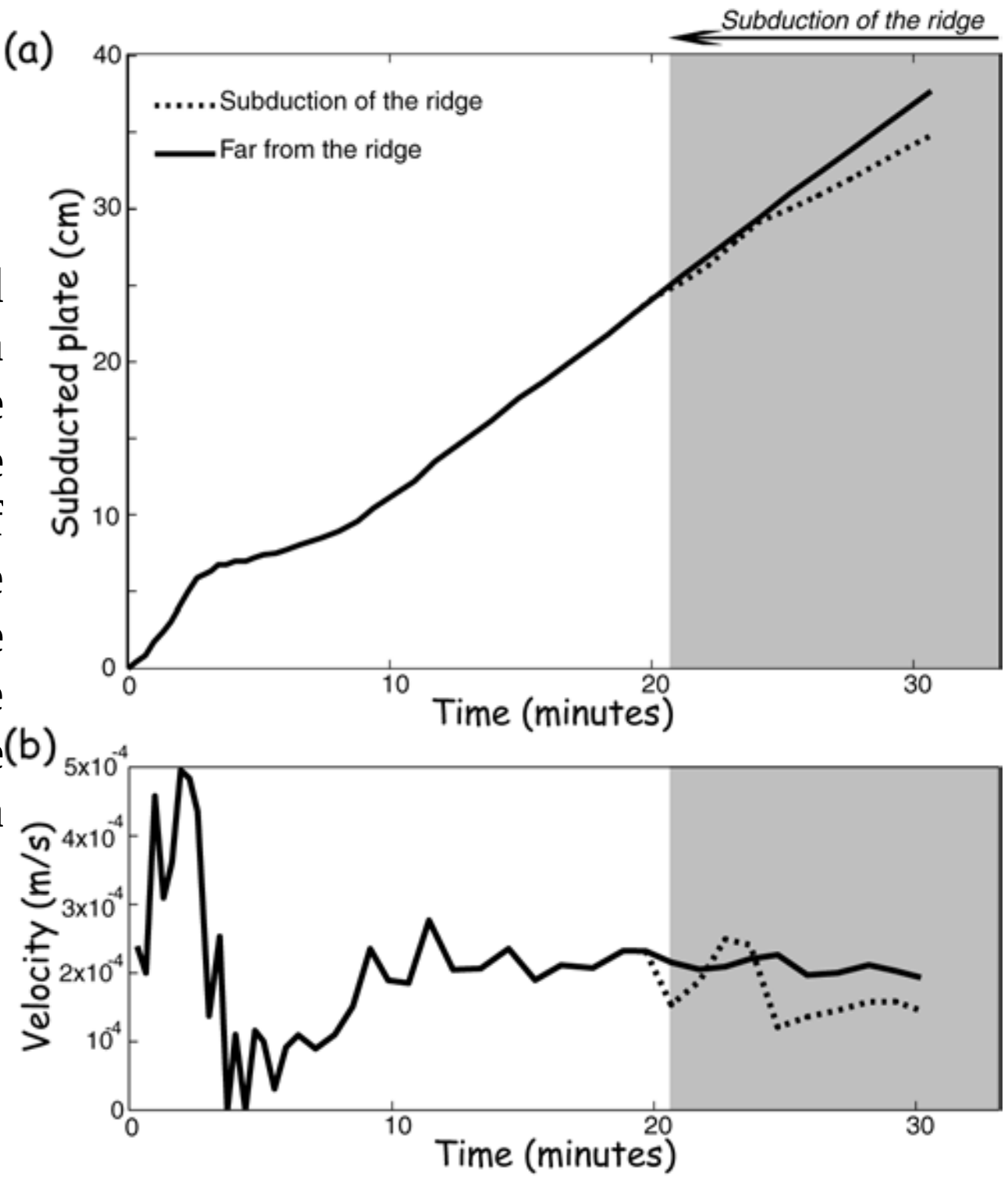

**Figure 11: (a) Amount of subducted plate and (b) velocity of subduction vs. time in experiment 8. The solid line corresponds to the subduction of the plate far from the ridge. The dotted line corresponds to the subduction of the ridge. The velocity of subduction of the ridge is only slightly smaller than that of the dense oceanic plate, because the motion of the rest of the plate forces ridge subduction. The geometry of the trench changes slowly as a consequence of the subduction of the ridge**

Although the ridge affects the geometry of the subducting slab, resulting in the formation of the cusp trench, we do not observe any significant change in the dip of the slab. The ridge volume, in spite the buoyant region models the equivalent of a 200 km-large and 1400 km-long ridge, is not large enough to significantly alter the dip of the slab, that preserves the same value close to 60° as before the entrance of the ridge into subduction. In fact, the ridge is pulled down by the rest of the plate, evidencing that predictions of the local behaviour of the slab using 1D (local) buoyancy analyses are not appropriate, as they do not take into account the rigidity of the lithospheric plate. It suggests that a 200 km-wide oceanic ridge perpendicular to the trench is not a sufficiently large perturbation to sharply alter the main characteristics of the process of subduction at depth.

### 3.5 Subduction of an oceanic plateau (experiments 9 and 10; Figure 2e)

Experiments 9 and 10 simulate the subduction of an oceanic plateau. The modelled plateau is parallel to the trench, 15 cm-long and 5 cm-large (Table 2 and Figure 12). Experiments 9 and 10 have been done with similar parameters, except that the plateau buoyancy is smaller in experiment 10 (relative plateau elevation of 1250 m vs. 1800 m for experiment 9) and its position within the dense plate closer to the boundary where subduction starts in experiment 10. It is possible to see in experiment 10 the end of the perturbation linked with the subduction of the light plateau, which is not possible in experiment 9 because the geometry of the slab remains perturbed by the plateau much time after it has been subducted (see below). The corresponding buoyant anomaly would approximately correspond in nature to a 1000 km per 350 km plateau , i.e. possibly somewhat similar to the Inca plateau plus the tip of the Nazca ridge that would control, according to Gutscher *et al.* (2000), the present-day geometry of the Peruvian flat slab segment. Note that examples of oceanic plateaus with a significantly larger surface and buoyancy exist. The surface of the Ontong Java and Kerguelen plateaus reach $2x10^6$ km$^2$ (vs. $0.35x10^6$ km$^2$ for the plateau modelled here), and the average topography of the Kerguelen plateau above the neighbouring dense oceanic floor is ~3000 meters.

The beginning of subduction and the appearance of the steady-state regime of subduction are similar to that of the other experiments. As soon as the plateau reaches the trench, the speed of subduction in the centre of the experiment diminishes. In the centre of the plateau, it lowers to 40% of the steady-state velocity of subduction of the dense oceanic plate, while on the borders of the plateau, the velocity decrease is smaller (Figure 13). The speed of subduction far from the plateau also decreases, although moderately. The trench takes rapidly an arched geometry to accommodate the different velocities of subduction (Figure 12).

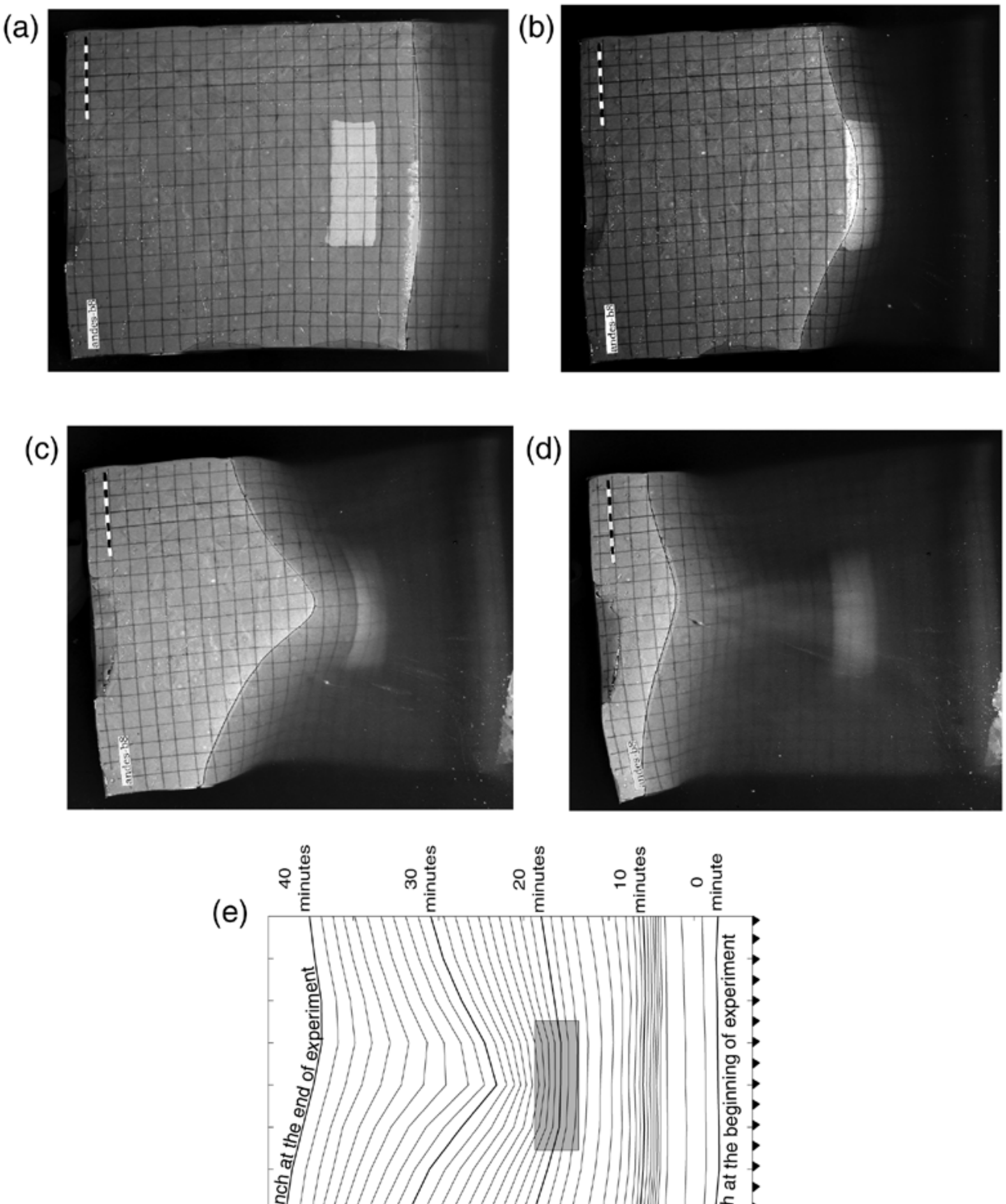

**Figure 12: Upper view of experiment 10. The 2 cm per 2 cm grid of passive markers gives the scale. (a) After 13mn11s of experiment, the plateau is approaching the trench. (b) After 21mn54s, the plateau is subducting. (c) After 30mn33s, the rate of subduction in the centre of the plate remains small as a result of the subduction of the plateau, and the geometry of the trench has been strongly modified. The dip of the upper part of the slab is smaller in the centre of experiment. (d) After 38mn16s of experiment. The velocity of subduction in the centre of the plate has accelerated, and the ridge is again nearly linear. The plateau is still slowly sinking into the asthenosphere, and the dip of the slab above it is close to 15° (e) Minute per minute evolution of the trench geometry in experiment 10. The grey rectangle corresponds to the position of the light plateau.**

The dip of the slab far from the plateau conserves approximately its previous 60° value. In contrast, in the centre of the experiment where the plateau is subducting, the geometry of the slab changes and resembles that described in experiment 4: The buoyant plateau sinks more slowly within the mantle. Meanwhile, the lower part of the slab below the plateau continues to fall rapidly, and the dip of the slab below the plateau increases to become nearly vertical. In experiment 10, the entire plateau is subducted after 24 minutes. However, the newly established subduction velocities (small in the centre of the plate and larger on both sides) remain constant during roughly 5 more minutes, during which the arched geometry of the trench amplifies. The superficial part of the slab above the plateau in the centre of the experiment becomes longer and flatter. After 30 minutes of experiment, the length of the slab presenting a small dip angle above the oceanic plateau is roughly 5 cm (equivalent to 340 km in Nature).

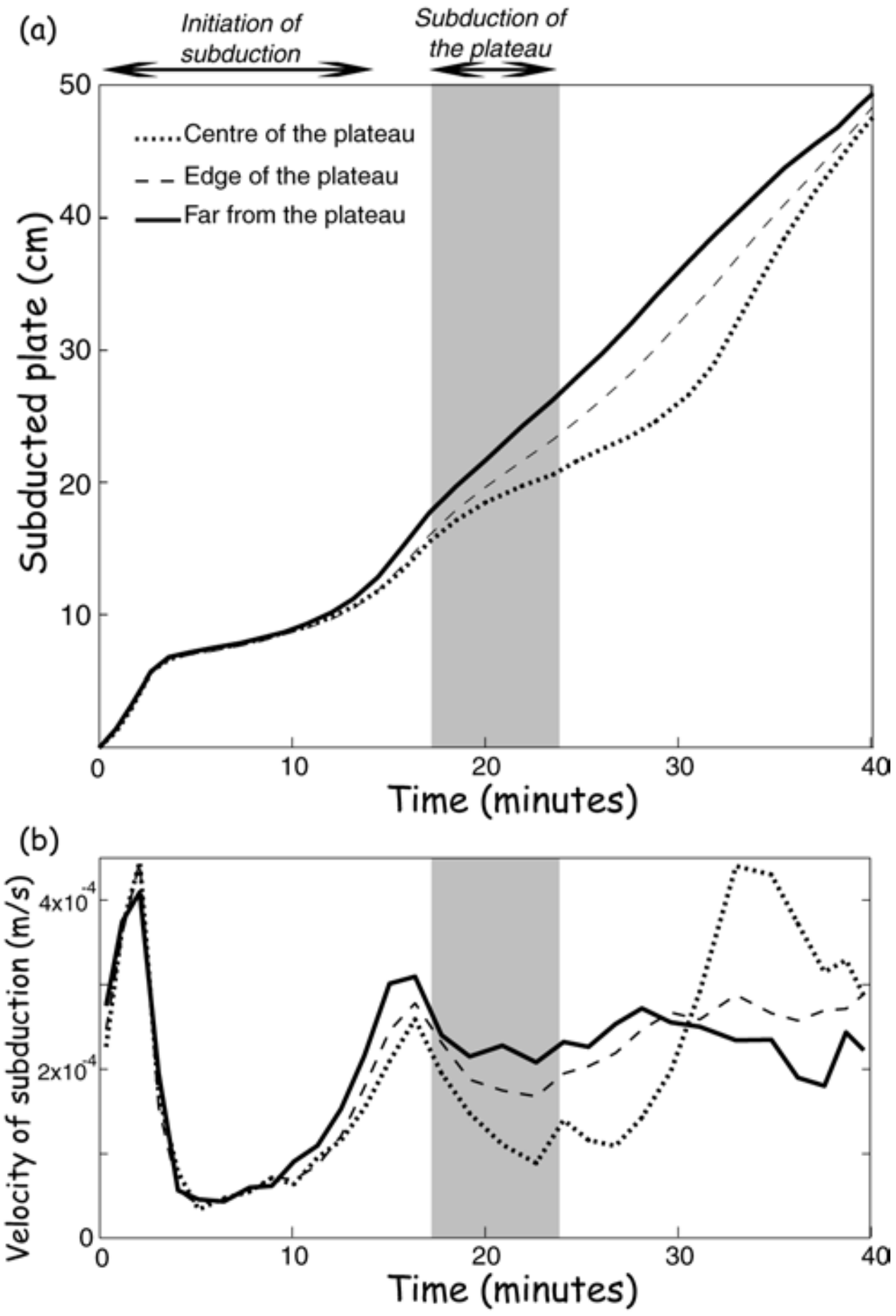

**Figure 13: Amount of subducted plate vs. time in experiment 10. The solid line corresponds to the subduction of the plate far from the plateau, the dotted line to the subduction of the centre of the plateau, and the dashed line to the subduction of the boundaries of the plateau.**

At that time, the velocity of subduction of the central part of the plate increases again. The arched form of the trench begins to attenuate (Fig. 12e). Meanwhile, the plateau sinks slowly within the mantle, and the dip of the slab above the plateau preserves a low value close to 15°. The plateau falls slowly within the mantle, because of its large buoyancy, and because the upper mantle lying below it has to move far backward to bypass the lateral parts of the slab that already attained the 660 km-discontinuity.

After 35 minutes of experiment, the dip of the slab in the centre of the experiment increases again. Only after 40 minutes of experiment, the dip of the subduction zone becomes again roughly the same in the middle and on the lateral boundaries of the experiment. Note that the perturbation resulting from the subduction of the plateau on the geometry of the trench lasted roughly 20 minutes in experiment 10 (Figure 13), despite the entire plateau subducted in only 6 minutes. Extrapolated to natural conditions, these periods could correspond to about 30 and 10 Ma, respectively (Table 1). Note, however, that this extrapolation only gives an order of magnitude of the corresponding natural time, since the scaling depends on many natural parameters that are not precisely known, and because of the numerous approximations described above.

The effect of the subduction of a light plateau with the geometry adopted in experiments 9 and 10 is clearly much more noticeable than that of the subduction of an oceanic ridge perpendicular to the trench. The dip of the slab above the plateau diminishes to 15°, as a consequence of the low subduction of the plateau and of the three-dimensional geometry of the slab that continue to subduct rapidly on both sides of the plateau. We also observe that the geometry of the slab remains affected by the subduction of the plateau several Ma after it has been subducted, because the descent of the plateau within the upper mantle is particularly slow.

## 4. Discussion

*Velocity of subduction*

The simple experiments described above show the effect on subduction of density variations that may result from the presence of oceanic plateaus or ridges in the subducting oceanic plate. In those experiments, the main force driving the process of subduction is the slab pull arising from the negative buoyancy of lithospheric plates. Resistive forces that dissipate the potential energy are, on the one hand those that are necessary to fold the lithospheric plate at trench and at the bottom of the tank, and on the other hand forces related to the slab-asthenosphere interaction. Subduction velocity $U$ results from the equilibrium between driving and resistive forces (Conrad & Hager, 1999; Funiciello *et al.*, 2003; Bellahsen *et al.*, 2005):

$$U = (C_S F_{SP} - C_F \tau l) / (2C_L \eta_L (h/r)^3 + 3 \eta_M C_M) \qquad (1)$$

where and $C_S$, $C_F$, $C_L$ and $C_M$ are constants respectively for slab pull, shear at subduction fault, lithospheric bending and mantle viscous dissipation, respectively; $F_{SP}$ is the slab pull force, $\tau$ is the shear stress generated at the subduction fault, $l$ is the length of the subduction fault, $\eta_L$ and $\eta_M$ are the lithosphere and upper mantle viscosities, $h$ is the thickness of the lithosphere, and $r$ is the radius of curvature of the slab.

In these experiments, the subduction fault is not modelled. Moreover, the viscosity ratio between the silicone plate and the underlying liquid being approximately $10^4$, we estimate that most of the energy is dissipated in folding the silicone plate (Bellahsen *et al.*, 2005). Then, equation (1) simplifies to

$$U = D F_{SP} r^3 / (2 \eta_L h^3) \qquad (2),$$

where $D = C_s/C_l$ is a dimensionless constant. We check that our experiments obey this relationship, the $D$ constant being roughly 2.5 (Table 3). Depending on the experiment, the observed $D$ value varies from 1.8 to 3.4.

We note that the effect on the velocity of subduction of the decreasing buoyancy resulting from the entrance of the ridge into the subduction zone is amplified by the increase of the dip of the slab which, in turn, causes a diminution of the radius of curvature (see equation 2). In experiment 3, for example, subduction nearly stops although the slab is still pulling the subducting plate. In experiment 4, during the subduction of the light ridge, the slab pull force is 44% of that prevailing during the subduction of the dense oceanic plate. Meanwhile, the velocity of subduction has fallen to 12% of the dense oceanic plate velocity of subduction.

*Slab geometry*

Analogue experiments suggest that denser plates are prone to retreat rapidly, favouring a larger radius of curvature of the slab in the subduction zone, which in turn favours a shallowing of the corresponding slabs. This observation contradicts the popular idea, typified by the Mariana example, that old plates should subduct vertically into the mantle. Compilation of the worldwide slab geometries shows, indeed, that old slabs are not steeper than young ones (e.g. Jarrard, 1986; King, 2001; Heuret & Lallemand, 2005). Note that although the density contrast influences the dip of the slab in our experiments, many other parameters whose effect has not been tested in this series of experiments, such as ratios between mantle and lithosphere viscosity, between mantle and lithosphere thickness, or the width of the subducting plate, are involved in the geometry of the retreating slab (Funiciello et al, 2004; Bellahsen et al., 2005). The dip of the slab results from the balance between the roll-back induced mantle flow force on the one hand, and the bending force on the other hand. The roll-back flow force is difficult to quantify in these experiments. Analyses of the energetic balance of this kind of experiments have been done in previous papers (Schellart, 2004; Bellahsen et al., 2005). Schellart (2004) estimates the energy dissipated by the viscous deformation of the mantle substracting the energy dissipated in bending the plate from the loss of potential energy. He points out that this estimation cannot be done precisely, in particular because it is difficult to have a correct estimation of the bending force in experiments. Estimation of this energy using analytical calculations is not straightforward given the 3D configuration of mantle flow (Funiciello et al., 2005). We plan to develop this analysis, which is beyond the scope of this paper, using a more complete set of experiments built specifically for constraining that question.

Analogue experiments show that density contrasts within the subducting plate and resulting from the presence of a thick oceanic crust may, in some cases, significantly alter the characteristics of subduction. We note, however, that the perturbation must imply large volumes of subducting plate to be able to modify the process of subduction. For instance, experiment 8 simulates the subduction of a 200 km-large oceanic ridge perpendicular to the trench buoyancy, and we observe that the entrance at trench of the ridge does not modify the dip of the slab. The volume of subducting buoyant lithosphere is too small, and the light ridge is essentially forced into subduction by the rest of the dense oceanic plate. The rigidity of the lithosphere opposes the deformation of the subducting plate, and although the trench takes an arched shape as a result of the subduction of the ridge, this shape only appears slowly and it is much smaller than the amount of subducted oceanic ridge.

In experiment 8, the subduction of the light ridge perpendicular to the trench results in a smaller velocity of subduction of the buoyant part of the plate (~0.15mm/s vs. 0.22 mm/s for the subduction of the homogeneous dense plate, see Figure 11). The ridge also influences laterally the velocity of subduction of the dense plate, at a distance depending on the viscous rigidity of the plate. In that experiment, we observe that plate subduction is slower at a distance from the ridge smaller than 8 cm, that distance increasing with time, i.e. with the amount of subducted ridge that increases the arched shape of the trench. This observation evidences that the ridge velocity results from the balance between the buoyancy of the ridge and that of a segment of the dense slab whose width ($W_{plr}$) is function of the plate rigidity. In the series of experiments presented above, an order of magnitude of the corresponding width can be estimated to ~20 cm, which would correspond in nature for a 50 Ma-old plate to approximately 1300 km.

This simple observation may explain why, in nature, the subduction of a light ridge does not necessarily modify the dip of the slab. In fact, those experiments show that 1D buoyancy (local isostatic) considerations are not appropriate to predict slab attitude, as already noted by van Hunen *et al.* (2002). The effect of the subduction of buoyant material and its tendency to flat upward the slab is counter balanced by the negative pull of the surrounding lithosphere and, hence, its effect depends on the plate strength and on the coupling between the ridge and the slab. Experiments suggest that a 200 km-large ridge entering a subduction zone perpendicularly to the trench may not alter significantly the dynamics of the slab, because the buoyancy contrast associated to that ridge should be averaged over a width that depends on the rigidity of the plate, comparison with analogue experiments suggesting that the corresponding width is larger than 1000 km for a 50 My-old plate. A ridge such as the Juan Fernandez ridge, for instance, only corresponds to an average extra sea-floor elevation $S$ of roughly $10^8$ $m^2$ per meter of ridge (see previous chapters). This topography can be related to an increase of the average buoyancy of the subducting oceanic plate :

$$\delta\rho = S\,(\rho_{ast} - \rho_w) / (W_{plr}\,H),$$

where $\rho_{ast}$ is the density of asthenosphere, $\rho_w$ that of water, $H$ is the thickness of the plate and $W_{plr}$ is the width that depends on plate rigidity over which we average the buoyancy of the plate. For $W_{plr}$ = 1300 km (see above), the average buoyancy variation resulting from the present-day Juan Fernandez ridge is only +2 kg/m$^3$, which is small compared to the normal buoyancy of a 50 My-old plate (-35 kg/m$^3$). Experiments suggest that the subduction of such a ridge should not alter significantly the process of subduction. In fact, the change in the Nazca slab geometry that coincides with the subduction of that ridge could result from the Neogene subduction of a ridge segment parallel to the trench (Yañez et al., 2001).

Experiments show, indeed, that the process of subduction is much more sensitive to the subduction of a ridge when the ridge is parallel to the trench (e.g. experiment 4), because a wider area of the subduction zone is simultaneously affected by the arrival of buoyant material at trench. In the presented experiments, the arrival of large amounts of buoyant material at the subduction zone results in a diminution of the speed of subduction and of trench retreat and, as a consequence, in an increase of the dip of the denser lower part of the slab. As a matter of fact, the dense lower part of the slab pursues its descent within the asthenosphere at a relatively high speed. One can imagine that in some cases, this situation may favour slab break-off, or the development of Rayleigh-Taylor instabilities (e.g. Regard *et al.*, 2003). The mechanical parameters (Newtonian viscous rheology) adopted in this set of experiments do not favour the occurrence of this phenomenon that needs localized deformation of the plate, and we cannot exclude that it could be observed in experiments with different boundary conditions and slab rheologies (e.g. non Newtonian).

Above the subduction of a light ridge parallel to the trench, the dip of the slab is not smaller than that of the subduction of a dense homogeneous plate (Fig. 8). Above oceanic plateaus (e.g. experiments 9 and 10), in contrast, the dip of the slab is significantly smaller than the dip of a dense homogeneous oceanic plate. This is a consequence of the 3D configuration of these experiments. The dense parts of the slab on both sides of the oceanic plateau sink more rapidly within the asthenosphere than the plateau, resulting in a very gently dipping slab above the plateau. Below the buoyant plateau, in contrast, the dip of the subducting plate is large to accommodate the different velocity of subduction between the heavy and light parts of the plate. This observation highlights the importance of taking into account the 3D phenomena occurring during subduction in experimental models. Experiments 9 and 10 also show that the perturbation of the slab geometry maintains for a long time after the entire plateau has been incorporated into the subduction zone. It may explain why variations in slab geometries are not always clearly associated with present-day subducting light ridges visible in oceanic seafloors.

### *Flat slabs, and comparison with the Andean subduction zone*

Finally, our results also point out the complexities to obtain flat subductions like for instance those described below Central Peru (3°-15S) and North-Central Chile (27°-33°S). In natural flat slab segments such as in Central Chile, the oceanic lithosphere passes just below the continental plate, and pursues horizontally at a constant depth of approximately 120 km on several hundreds of kilometres (e.g. Cahill & Isacks, 1992; Pardo *et al.*, 2002). On that distance, the two lithospheric plates touch themselves. Earthquakes occurring at plate interfaces, and the location of compressive earthquakes occurring within the continental lithosphere just above the place where the slab detaches from the upper plate to descend into the asthenosphere, both evidence the friction existing between the plates along the entire flat slab segment. Those observations show that the top of the oceanic plate is forced to maintain at a depth of 120 km by the overlying continental lithosphere. Then, flat slab segments probably result both from the motion of the overriding plate, and from the subduction of buoyant anomalies sufficiently large to perturb the local dynamics of the slab. The combined effect of subduction of aseismic ridges and motion of the overriding plate will be studied in detail in further experiments.

The simple experiments modelling the subduction of oceanic plateaus presented above, however, show that the process of subduction depends on the average buoyancy of a wide lithosphere segment parallel to the trench (~1300 km for a 50 Ma-old plate). Modest ridges subducting perpendicular to the trench, such as the present-day Juan Fernandez ridge, do not decrease significantly the average buoyancy of the plate (see above), and experiments suggest they should not explain the presence of flat-slab segments. In central Chile, however, Yañez *et al.* (2001) argue that the Juan Fernandez ridge is associated to the flat slab segment. Comparison with analogue experiments suggest that this flat segment may have appeared because the ridge that subducted between 14 and 10 Myr ago was almost parallel to the trench. The Central Chile flat-slab segment could be, in fact, a consequence of this episode of ridge-parallel to the trench subduction. Similarly, below Peru, the other present-day flat-slab segment may have resulted from the subduction of both the Nazca ridge and the Inca plateau (Gutscher *et al.*, 2000), i.e. from the subduction of some 1300 km parallel to the trench buoyant oceanic lithosphere. In both case, the subduction of large volumes of buoyant lithosphere parallel to the trench would have helped in the formation of the flat slab segment.

Note also that the observation of seismicity associated to the subduction of the Nazca slab suggests that the dip of the slab is larger below flat slab segments, as observed in analogue experiments below the subduction of light plateaus. As a matter of fact, the deep (>600 km) seismicity below South America is roughly parallel to the trench, independently of the shallow geometry of the slab, showing that the dip of the slab is greater at large depths if smaller near the surface (e.g. Cahill & Isacks, 1992). Tomographic images (Gutscher *et al.*, 2000) also suggest that the slab at depths between 200 and 600 km is steeper below flat slab segments.

## 5. Conclusions

We performed analogue experiments to study the subduction of an oceanic lithosphere containing buoyant ridges or plateaus. In this set of experiments, the initiation of the subduction of the oceanic plate is followed by a steady-state “mode 1” (Bellahsen et al., 2005) regime of subduction during which the trench retreats, the velocity of subduction and the dip of the slab remaining constant. This steady-state regime is highly reproducible, and it has been observed in all the experiments. We observe how density heterogeneities within the plate perturb the steady-state regime of subduction.

Subduction of buoyant segments of oceanic lithosphere resulting in a diminution of the slab-pull force may decrease the velocity of subduction. Experiments show, however, that the subducting buoyant segment must be large enough to alter the dynamics of subduction. Narrow oceanic ridges perpendicular to the trench are essentially pulled down into subduction by the rest of the plate, and do not modify markedly the dip of the slab, evidencing that 1D buoyancy (local isostatic) considerations are not appropriate to predict slab attitude (van Hunen *et al.*, 2002).

In case the buoyant segment is large enough to resist the pull exerted by the rest of the plate, it will tend to sink more slowly into the upper mantle. The dip of the slab below it will increase to accommodate this velocity difference. Above buoyant subducting plateaus, the dip of the slab drops significantly. This has been observed, for instance, in experiments 9, where the dip of the slab is only 15° above the subducted plateau, while it is close to 50° where the dense oceanic plate is subducting. We do not reproduce flat slab subductions in this experimental set, because they probably also result from the interaction with the overlying continent, and because we do not take into account the upper plate in these experiments. Experiments suggest, however, that large heterogeneities such as oceanic plateaus or ridges parallel to the trench entering the subduction could affect the slab geometry and result in the appearance of a flat slab segment.

**Acknowledgments:** Analogue experiments have been performed in the Laboratory of Experimental Tectonics of University “Roma TRE”. We thank the four anonymous reviewers for their helpful comments.